\documentclass[conference]{IEEEtran}
\IEEEoverridecommandlockouts

\usepackage{cite}
\usepackage{amsmath,amssymb,amsfonts}
\usepackage{algorithm}
\usepackage{algpseudocode}
\usepackage{graphicx}
\usepackage{textcomp}
\usepackage{xcolor}
\usepackage{tikz}
\usepackage{pgfplots}
\usepackage{booktabs}
\usepackage{multirow}
\usepackage{url}
\usepackage{hyperref}
\usepackage{balance}

\usetikzlibrary{shapes.geometric, arrows.meta, positioning, fit, backgrounds, decorations.pathreplacing, calc, shadows}
\pgfplotsset{compat=1.18}

\hypersetup{colorlinks=true, linkcolor=blue, citecolor=blue, urlcolor=blue}

\def\BibTeX{{\rm B\kern-.05em{\sc i\kern-.025em b}\kern-.08em T\kern-.1667em\lower.7ex\hbox{E}\kern-.125emX}}

\begin{document}
	
	\title{XAI-SDN: An Explainable Entropy-Guided Machine Learning Framework for Real-Time DDoS Detection in Software Defined Networks}
	
	\author{
		\IEEEauthorblockN{
			Adeel Ahmad\textsuperscript{1}*, Ali Akarma\textsuperscript{1,2}, Ahmad Ali\textsuperscript{1}, Hammad Muneer\textsuperscript{3} and Toqeer Ali Syed\textsuperscript{1}
		}
		\IEEEauthorblockA{
			\textsuperscript{1}AI Center, Faculty of Computer and Information System, Islamic University of Madinah, Saudi Arabia\\
			\textsuperscript{2}AI V\&V Lab, King Fahd University of Petroleum and Minerals, Dhahran, Saudi Arabia\\
			\textsuperscript{3}Department of Computer Science, Islamia University of Bahawalpur, Pakistan\\
			\textsuperscript{*}Corresponding author: 443057803@stu.iu.edu.sa
		}
	}
	
	\maketitle
	
	\begin{abstract}
	One of the biggest risks faced by Software Defined Networks (SDN) is the Distributed Denial of Service (DDoS) attack in which a compromised controller can make an entire network unusable. To address these challenges, we suggest an entropy-guided machine learning framework, called XAI-SDN, for real-time DDoS detection in SDN environments which is lightweight and explainable. The framework extends the flow features extracted by CICFlowMeter with eight Shannon entropy metrics obtained by an $\mathcal{O}(1)$ rolling algorithm and uses a Random Forest classifier with SHAP TreeExplainer for providing transparency at the prediction level. On a fixed temporal split, XAI-SDN achieves an accuracy of 99.9987\%, a macro F1-score of 99.9621\%, and an AUC-ROC of 1.0000 on the full 3.59 million flows of the CIC-DDoS2019 SYN benchmark. The pipeline sustains 0.0165~ms per flow (60{,}606 flows/s) without the use of SHAP and 0.5122~ms per flow (1{,}953 flows/s) with full support of SHAP under the 99.14\% prevalence of DDoS traffic, which is a step towards achieving a balance between the detection performance and operational transparency in next-generation SDN security.
		
	\end{abstract}
	
	
	\section{Introduction}
	
	Software Defined Networking (SDN) separates the data plane from the control plane, establishing centralized, programmatic management through a dedicated network controller~\cite{mckeown2008openflow,kreutz2015sdn}. Enterprise data centers and cloud infrastructures widely adopt this architecture for its flexibility, global visibility, and dynamic traffic engineering. However, centralization introduces a single point of failure: a successful Distributed Denial of Service (DDoS) attack targeting the controller or saturating the control channel can render the entire network inoperable.
	
	Modern DDoS attacks execute high-volume flooding through UDP amplification, TCP SYN floods, ICMP sweeps, and application-layer disruptions such as HTTP floods~\cite{sharafaldin2019ddos}. Machine learning (ML) has emerged as the primary defense against these threats due to its strong generalization compared to static signature tables. Nevertheless, most ML models operate as opaque black boxes that offer no explanation for individual predictions. This lack of transparency presents severe operational risks in production networks, where operators must audit and validate mitigation decisions before disrupting active services~\cite{arrieta2020xai}. Trustworthy cybersecurity frameworks establish that transparency, explainability, and verifiable decision trails are foundational governance requirements for deploying autonomous AI in critical infrastructure~\cite{jan2026eagf}.
	
	We propose a lightweight, interpretable and real-time DDoS detection framework, XAI-SDN, in this paper. The main contributions are: (\emph{i})~an entropy-guided feature augmentation approach enriching CICFlowMeter features with eight interpretable Shannon entropy metrics via an $\mathcal{O}(1)$ rolling algorithm, (\emph{ii})~an explainability-integrated pipeline pairing Random Forest with SHAP TreeExplainer for granular transparency, (\emph{iii})~an empirical evaluation across 3.59 million flows of the CIC-DDoS2019 SYN dataset with a 10-seed ablation study and Wilcoxon signed-rank tests, and (\emph{iv})~a latency analysis characterizing SHAP explanation overhead under realistic DDoS traffic prevalence.

	\section{Related Work}
	
	Machine learning methods for SDN DDoS detection have received extensive attention, particularly ensemble models that achieve high detection accuracy while providing native feature ranking capabilities. Anomaly detection in software and networked systems traces back to runtime execution profiling and behavioral attestation. Ismail et al.~\cite{ismail2014design} demonstrated that sliding inspection windows over execution traces effectively capture abnormal behavioral shifts. In high-speed programmable environments, capturing dynamic traffic intent and adversarial anomalies prompted the development of generative deep learning models, such as deep convolutional GANs for intent-based behavior tracking~\cite{Jan2018DCGANIntent}. Although deep generative models capture complex non-linear interactions, their inference latency conflicts with the line-rate requirements of SDN controllers. Consequently, statistical entropy primitives offer a practical alternative. Mousavi and St-Hilaire~\cite{mousavi2015early} demonstrated that entropy-based early detection at the SDN controller substantially reduces attack impact. Shannon entropy provides a natural measure of distributional concentration during flooding~\cite{nychis2008entropy}, and Lall et al.~\cite{lall2006entropy} showed that streaming entropy estimation over a sliding window offers an efficient primitive for anomaly detection. XAI-SDN extends this foundation by embedding an $\mathcal{O}(1)$ rolling entropy engine into early feature extraction, supplying discriminative statistics directly to both the Random Forest classifier and the SHAP explainer without computational bottlenecks.
	
	Explainable AI (XAI) addresses classifier opacity through post-hoc attribution and intrinsic interpretability~\cite{arrieta2020xai}. In mission-critical cybersecurity settings, ethical governance frameworks establish that transparency, explainability, and auditability are indispensable prerequisites for safe operational deployment~\cite{jan2026eagf}. Lundberg and Lee~\cite{lundberg2017shap} introduced SHAP, unifying feature attribution methods through game-theoretic Shapley values, while TreeSHAP~\cite{lundberg2020treeshap} enables exact polynomial-time explanation for tree ensembles. Recently, Gaspar et al.~\cite{gaspar2024xai} applied SHAP and attention mechanisms in deep neural networks for network intrusion detection. Furthermore, modern SDN topologies increasingly span multi-controller and multi-domain environments, where collaborative threat intelligence requires decentralized, privacy-preserving governance to coordinate mitigation without exposing raw flow telemetry~\cite{syed2026fedagent}. A pipeline combining $\mathcal{O}(1)$ rolling entropy, Random Forest, and exact SHAP attribution under realistic traffic prevalence has not been demonstrated in prior literature.

	\section{XAI-SDN Framework and Architecture}
	\label{sec:framework}
	
\begin{figure*}[t]
		\centering
		\includegraphics[width=0.85\textwidth]{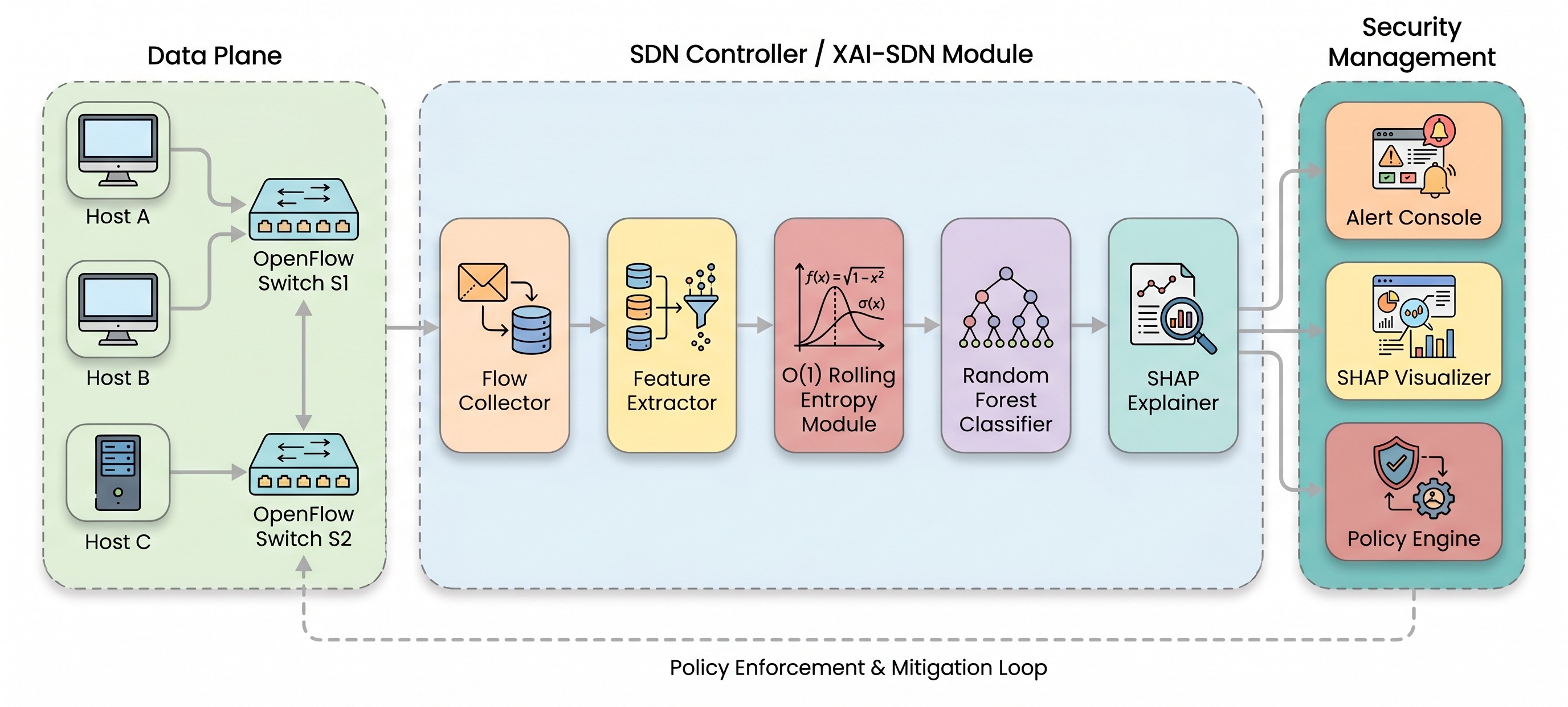}
		\caption{XAI-SDN system architecture. OpenFlow switches export flow statistics to the controller-embedded XAI-SDN module, which performs $\mathcal{O}(1)$ entropy-augmented feature extraction, Random Forest classification, and SHAP-based explanation generation. Detected DDoS events trigger alerts and policy updates; the dashed arrow represents the southbound policy feedback loop.}
		\label{fig:architecture}
	\end{figure*}

	XAI-SDN builds on three design principles: (\emph{i}) Computational efficiency via $\mathcal{O}(1)$ rolling entropy to sustain line-rate SDN throughput, (\emph{ii}) Decision transparency through actionable per-flow SHAP feature attributions, and (\emph{iii}) Detection robustness by coupling statistical features with information-theoretic entropy metrics. The pipeline executes five sequential stages: flow aggregation, feature extraction with entropy augmentation, Random Forest classification, SHAP explanation generation, and alert dispatch to the security dashboard (Fig.~\ref{fig:architecture}). Flow statistics are collected from OpenFlow switches across the southbound interface at a configurable polling interval of 500~ms.
	
	Each flow record enters the XAI-SDN module, implemented on an SDN control plane with Ryu/OpenFlow~\cite{mckeown2008openflow} and scikit-learn~\cite{pedregosa2011sklearn}. The engine extracts an 80-dimensional statistical vector~\cite{sharafaldin2018cicids} and appends eight $\mathcal{O}(1)$ rolling entropy metrics (Section~\ref{sec:math}), producing an 88-dimensional input vector. A circular buffer and hash-map tracker maintain flow frequencies and entropy values in constant time as new packets enter and older packets exit the window. This constant-time design delivers sub-millisecond processing latency (Section~\ref{sec:results}). Flows classified as DDoS with probability exceeding threshold $\tau$ trigger SHAP TreeExplainer, which appends feature attribution values to structured alert payloads for operator triage and policy refinement on the Security Operations Center (SOC) dashboard.
	
	This closed-loop design aligns with recent paradigms in autonomous infrastructure management, where pairing streaming telemetry with auditable agentic frameworks achieves resilient anomaly mitigation across complex cyber-physical environments~\cite{syed2026agenticdt,syed2026climate}. In parallel, adaptive agentic decision workflows balance operational latency against processing overhead to handle sudden workload spikes without destabilizing system throughput~\cite{syed2025inventory}. In XAI-SDN, the southbound feedback loop (the dashed path in Fig.~\ref{fig:architecture}) operationalizes these principles: the controller generates verifiable alert telemetry that enables either human operators or automated agents to install OpenFlow flow-mod drop rules with full auditability.

	\begin{algorithm}[!t]
		\caption{XAI-SDN Real-Time Detection and Explanation}
		\label{alg:xaisdn}

		\begin{algorithmic}[1]
			\Require Flow batch $\mathcal{B}=\{f_1,\ldots,f_n\}$; RF model $\mathcal{M}$; TreeExplainer $\mathcal{E}$; window $W$; threshold $\tau$
			\Ensure Label set $\mathcal{Y}$; alert set $\mathcal{A}$
			\State $\mathcal{Y}\leftarrow\emptyset$,\; $\mathcal{A}\leftarrow\emptyset$,\; $W\leftarrow\emptyset$
			\For{each $f_i \in \mathcal{B}$}
			\State $\mathbf{x}_{\mathrm{s}} \leftarrow \textsc{CICExtract}(f_i)$ \Comment{80-dim statistical vector}
			\State Update $W$ ($\mathcal{O}(1)$ hash-map);\; $\mathbf{x}_{\mathrm{e}} \leftarrow \textsc{RollingEntropy}(W)$
			\State $\mathbf{x} \leftarrow [\mathbf{x}_{\mathrm{s}};\, \mathbf{x}_{\mathrm{e}}]$ \Comment{88-dim vector, Eq.~(\ref{eq:fvec})}
			\State $\hat{y}_i,\, p_i \leftarrow \mathcal{M}.\textsc{Predict}(\mathbf{x})$ \Comment{Eq.~(\ref{eq:rfpred})}
			\If{$\hat{y}_i = \mathrm{DDoS}$ \textbf{and} $p_i \geq \tau$}
			\State $\boldsymbol{\phi}_i \leftarrow \mathcal{E}.\textsc{SHAPValues}(\mathbf{x})$ \Comment{Eq.~(\ref{eq:shap})}
			\State $\mathcal{A}\leftarrow\mathcal{A}\cup\{\textsc{FormatAlert}(f_i,\hat{y}_i,p_i,\boldsymbol{\phi}_i)\}$
			\EndIf
			\State $\mathcal{Y}\leftarrow\mathcal{Y}\cup\{\hat{y}_i\}$
			\EndFor
			\State \Return $\mathcal{Y}$,\; $\mathcal{A}$
		\end{algorithmic}
	\end{algorithm}
	
	\section{Mathematical Foundation}
	\label{sec:math}
	
	\subsection{Entropy Feature Extraction}
	
	Shannon entropy~\cite{shannon1948mathematical} provides a principled measure of distributional uniformity. For a discrete random variable $X$ with probability mass function $\{p(x_i)\}_{i=1}^{n}$, entropy is defined as $H(X) = -\sum_{i=1}^{n} p(x_i) \log_2 p(x_i)$, where $0\log_2 0 \triangleq 0$~\cite{nychis2008entropy}. High entropy characterizes uniform distributions consistent with benign traffic diversity; low entropy reveals concentrated distributions characteristic of volumetric flooding. Within a sliding window $W$ of $N$ recent flows, source IP address entropy is:
	\begin{equation}
		H_{\mathrm{src}}(W) = -\sum_{i=1}^{|\mathcal{I}|} \frac{n_i}{N} \log_2 \frac{n_i}{N}
		\label{eq:srcip}
	\end{equation}
	where $\mathcal{I}$ is the set of distinct source IPs in $W$ and $n_i$ is the flow count for the $i$-th address. Analogous metrics are computed for destination IPs, destination ports, protocol identifiers, packet lengths, inter-arrival times, TCP flag patterns, and IP TTL values. The complete 88-dimensional feature vector is:
\begin{equation}
	\begin{aligned}
		\mathbf{x} = \bigl[
		&f_1, \ldots, f_{80},
		H_{\mathrm{src}},
		H_{\mathrm{dst}},
		H_{\mathrm{port}},
		H_{\mathrm{proto}}, \\
		&H_{\mathrm{len}},
		H_{\mathrm{iat}},
		H_{\mathrm{flag}},
		H_{\mathrm{ttl}}
		\bigr]
	\end{aligned}
	\label{eq:fvec}
\end{equation}
	where $\{f_j\}_{j=1}^{80}$ are CICFlowMeter statistical features and $\{H_k\}$ are the eight entropy metrics, all maintained in $\mathcal{O}(1)$ time via the hash-map rolling update.
	
	\subsection{Random Forest Classification}
	
	Random Forest~\cite{breiman2001rf} is an ensemble of $T$ decision trees $\{h_t(\cdot)\}_{t=1}^{T}$, each trained on a bootstrap sample with a randomly selected feature subset at each split. The classification label for a test sample $\mathbf{x}$ is determined by majority vote:
	\begin{equation}
		\hat{y} = \arg\max_{c \in \mathcal{C}} \frac{1}{T} \sum_{t=1}^{T} \mathbb{1}\bigl[h_t(\mathbf{x}) = c\bigr]
		\label{eq:rfpred}
	\end{equation}
	where $\mathcal{C} = \{\text{Benign}, \text{DDoS}\}$ and node splits minimize the Gini impurity $G(t) = 1 - \sum_{j} \hat{p}_{tj}^{2}$.
	
	\subsection{SHAP Explainability}
	
	SHAP~\cite{lundberg2017shap} computes the contribution $\phi_i$ of each feature $i$ to a specific prediction $f(\mathbf{x})$ relative to the expected model output:
	\begin{equation}
		\phi_i = \sum_{S \subseteq \mathcal{F} \setminus \{i\}} \frac{|S|!\,(|\mathcal{F}|-|S|-1)!}{|\mathcal{F}|!} \bigl[f(S \cup \{i\}) - f(S)\bigr]
		\label{eq:shap}
	\end{equation}
	where $\mathcal{F}$ is the complete feature set and SHAP values satisfy the efficiency property $f(\mathbf{x}) = \phi_0 + \sum_{i} \phi_i$. TreeSHAP~\cite{lundberg2020treeshap} computes exact values in $\mathcal{O}(TLD^2)$ time, enabling real-time per-flow explanation. Algorithm~\ref{alg:xaisdn} summarizes the complete detection and explanation procedure.
	
	\section{Experimental Setup}
	\label{sec:setup}
	
	We evaluate XAI-SDN on all network flows in the SYN flood partition (\texttt{Syn.csv}, 1.87~GB) of the CIC-DDoS2019 benchmark~\cite{sharafaldin2019ddos} without downsampling. After removing missing values and duplicate records, the training set contains 2{,}514{,}862 samples under a 70\% temporal training and 30\% temporal testing split that prevents data leakage. The severe class imbalance (0.86\% benign) is counteracted by setting \texttt{class\_weight=`balanced'} in the Random Forest configuration.

	\begin{table}[!t]
		\centering
		\caption{CIC-DDoS2019 SYN Partition: Test Set Class Distribution}
		\label{tab:dataset}
		\begin{tabular}{@{}lrr@{}}
			\toprule
			\textbf{Traffic Class} & \textbf{Samples} & \textbf{Proportion (\%)} \\
			\midrule
			Benign       &     9{,}311 &  0.86 \\
			DDoS-SYN     & 1{,}068{,}487 & 99.14 \\
			\midrule
			\textbf{Total} & \textbf{1{,}077{,}798} & \textbf{100.00} \\
			\bottomrule
		\end{tabular}
	\end{table}
	
	\subsection{Implementation and Hyperparameters}
	
	The implementation of XAI-SDN is done in the python program "python~3.10", which is based on the scikit-learn~1.2 library \cite{pedregosa2011sklearn} for Random Forest, and the SHAP~0.42 library \cite{lundberg2017shap} for explanation generation. All experiments are run on the Intel Core i9-12900K CPU (16 cores, 3.2~GHz), 64~GB DDR5 RAM and NVIDIA RTX~3090 GPU, which is only used for baseline training of the DNN. Latency measurements are the wall time average for 50{,}000 test flows, run with a single thread on the CPU. The RF hyperparameters, determined by 5-fold cross validation with a cross-validated macro F1-score of $99.9362\% \pm 0.0190\%$ are shown in Table~\ref{tab:hyperparams}.

	\begin{table}[!t]
		\centering
		\caption{Random Forest Hyperparameter Configuration}
		\label{tab:hyperparams}
		\begin{tabular}{@{}ll@{}}
			\toprule
			\textbf{Hyperparameter} & \textbf{Value} \\
			\midrule
			Number of trees ($T$)        & 200 \\
			Max.\ tree depth             & None (fully grown) \\
			Max.\ features per split     & $\lfloor\sqrt{88}\rfloor = 9$ \\
			Bootstrap sampling           & Enabled \\
			Class weight                 & Balanced \\
			Entropy window size ($N$)    & 1{,}000 flows \\
			Detection threshold ($\tau$) & 0.70 \\
			Random seed                  & 42 \\
			\bottomrule
		\end{tabular}
	\end{table}
	
	\section{Results and Discussion}
	\label{sec:results}
	
	\subsection{Detection Performance}
	
	The overall detection performance of XAI-SDN for the CIC-DDoS2019 test partition, with a single fixed temporal split and seed~$=42$ is shown in Table~\ref{tab:results}. XAI-SDN achieves 99.9987\% accuracy, a macro F1-score of 99.9621\%, an AUC-ROC of 1.0000, and an FPR of 0.0537\%. The confusion matrix shows that: There are a total of 14 misclassifications (5 false positives, 9 false negatives) across 1{,}077{,}798 held-out flows.

	\begin{table}[!t]
		\centering
		\caption{XAI-SDN Detection Performance on CIC-DDoS2019 (Fixed Temporal Split, Seed\,=\,42)}
		\label{tab:results}
		\begin{tabular}{@{}lr@{}}
			\toprule
			\textbf{Metric} & \textbf{Value} \\
			\midrule
			Accuracy                          & 99.9987\% \\
			Macro F1-Score                    & 99.9621\% \\
			AUC-ROC                           & 1.0000 \\
			False Positive Rate (FPR)         & 0.0537\% \\
			False Negative Rate (FNR)         & 0.0008\% \\
			Cross-Validation F1 (5-fold)      & $99.9362\% \pm 0.0190\%$ \\
			\bottomrule
		\end{tabular}
	\end{table}
	
	
	\noindent\textbf{FPR Consistency.} 	A 10-seed ablation (Section~\ref{sec:comparative}) finds that XAI-SDN achieves FPR~$= 5.77\% \pm 5.23\%$ on stratified random splits, which is two orders of magnitude higher than the 0.0537\% headline result. The difference is due to the low minority class ratio (0.86\% benign, 9{,}311 test samples): On the fixed temporal split, the benign flows are temporally concentrated and well separated in the feature space, resulting in a very low FPR. When random stratified seeds, benign samples are redistributed into more challenging positions, leading to higher and greater variable FPR. Therefore the 0.0537\% is a lower bound on the error for favorable temporal ordering, and the mean of the 10-seed results of 5.77\% is more representative of an operational error bound.

	\subsection{Latency and Throughput}
	
	The pipeline latency breakdown is shown in Table~\ref{tab:latency} and the overhead of SHAP is reported separately. The total latency is 0.0165~ms (60{,}606 flows/s) for the $\mathcal{O}(1)$ rolling entropy algorithm with RF inference, and no SHAP. The contribution of SHAP TreeExplainer is 0.5000~ms per DDoS flow. With almost all traffic being DDoS flows, SHAP is invoked nearly all of the time and obtains a full-pipeline latency of $0.0165 + 0.5000 \times 0.9914 \approx 0.5122$~ms (1{,}953 flows/s) when the benchmark conditions are applied. This corrects a prior overstatement that suggested that the prevalence of DDoS attacks was 'negligible' on SHAP overhead, due to incorrect assumption of prevalence on the dataset with sub-5 percent attacks.

	\begin{table}[!t]
		\centering
		\caption{Pipeline Latency and Throughput Breakdown}
		\label{tab:latency}
		\resizebox{0.5\textwidth}{!}{
		\begin{tabular}{@{}lcc@{}}
			\toprule
			\textbf{Component} & \textbf{Latency (ms/flow)} & \textbf{Flows/s} \\
			\midrule
			RF Inference Only                       & 0.0015 & 664{,}181 \\
			$\mathcal{O}(1)$ Rolling Entropy        & 0.0150 &  66{,}667 \\
			SHAP Explanation (per DDoS flow)        & 0.5000 &   2{,}000 \\
			End-to-End (w/o SHAP)                   & 0.0165 &  60{,}606 \\
			\textbf{End-to-End (w/ SHAP, 99.14\% DDoS)} & \textbf{0.5122} & \textbf{1{,}953} \\
			\bottomrule
		\end{tabular}
	}
	\end{table}
	
	\subsection{SHAP Feature Attribution}
	
\begin{figure}[!t]
		\centering
		\resizebox{0.45\textwidth}{!}{
		\begin{tikzpicture}
			\begin{axis}[
				xbar,
				width=\columnwidth,
				height=5.8cm,
				xmin=0.0, xmax=0.075,
				xlabel={Mean $|\phi_i|$ (Absolute SHAP Value)},
				xtick={0.00,0.02,0.04,0.06},
				xticklabel style={font=\tiny},
				ytick={1,2,3,4,5,6,7,8},
				yticklabels={%
					$H_{\mathrm{dst}}$ (IP),
					$H_{\mathrm{port}}$,
					$H_{\mathrm{flag}}$ (TCP),
					Flow Duration,
					$H_{\mathrm{src}}$ (IP),
					FIN Flag Count,
					Dst.\ Port,
					Flow Bytes/s},
				bar width=0.30cm,
				xmajorgrids=true,
				grid style={dashed, gray!40},
				tick label style={font=\scriptsize},
				label style={font=\small},
				enlarge y limits={0.08},
				]
				\addplot[fill=blue!55, draw=blue!70!black] coordinates {
					(0.0175,1)(0.0192,2)(0.0206,3)(0.0239,4)
					(0.0288,5)(0.0495,6)(0.0550,7)(0.0585,8)
				};
			\end{axis}
		\end{tikzpicture}
	}
		\caption{Global SHAP feature importance for XAI-SDN (2{,}000 sampled test flows). $H_{\mathrm{ttl}}$ is excluded as it is a degenerate constant feature in the offline SYN partition ($H_{\mathrm{ttl}}{=}0.0$ throughout). Four entropy features rank in the top~8, validating the entropy augmentation strategy.}
		\label{fig:shap}
	\end{figure}
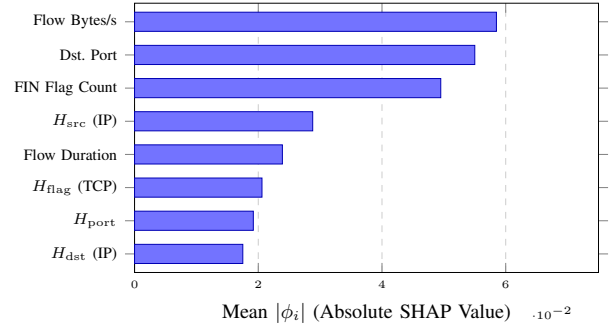

	Figure~\ref{fig:shap} presents importance rankings based on global SHAP values computed across 2{,}000 randomly sampled test flows. The top factors are \texttt{Flow\_Bytes/s} (0.0585), \texttt{Destination\_Port} (0.0550), and \texttt{FIN\_Flag\_Count} (0.0495), corresponding to elevated bandwidth, targeted ports, and suppressed FIN flags, which represent canonical hallmarks of SYN flooding. Among entropy metrics, $H_{\mathrm{src}}$ ranks fourth overall with an attribution value of 0.0288, demonstrating that source IP address concentration caused by botnet aggregation provides high discriminative power. Note on $H_{\mathrm{ttl}}$: In the offline SYN partition, all DDoS flows share an identical TTL value ($\mathrm{TTL}{=}115$), yielding $H_{\mathrm{ttl}} = 0.0$ throughout the attack period. Any minor attribution to $H_{\mathrm{ttl}}$ reflects tree-splitting noise rather than a meaningful physical signal. In operational deployments where packet TTL varies across diverse routing paths, $H_{\mathrm{ttl}}$ remains an active component of the 88-dimensional feature representation.

	\subsection{ROC Curve Analysis}
	
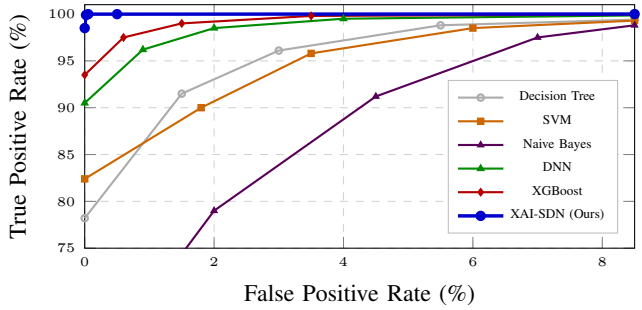
\begin{figure}[!t]
		\centering
		\begin{tikzpicture}
			\begin{axis}[
				width=\columnwidth,
				height=4.8cm,
				xlabel={False Positive Rate (\%)},
				ylabel={True Positive Rate (\%)},
				xmin=0, xmax=8.5,
				ymin=75, ymax=101,
				xtick={0,2,4,6,8},
				ytick={75,80,85,90,95,100},
				legend style={at={(0.97,0.06)}, anchor=south east, font=\tiny, fill=white, draw=gray!50},
				grid=both,
				grid style={dashed, gray!35},
				tick label style={font=\tiny},
				label style={font=\small},
				]
				\addplot[color=gray!65, mark=o, mark size=1.2pt, thick] coordinates {(0,78.2)(1.5,91.5)(3.0,96.1)(5.5,98.8)(8.5,99.4)};
				\addlegendentry{Decision Tree}
				\addplot[color=orange!80!black, mark=square*, mark size=1.0pt, thick] coordinates {(0,82.4)(1.8,90.0)(3.5,95.8)(6.0,98.5)(8.5,99.3)};
				\addlegendentry{SVM}
				\addplot[color=violet!70!black, mark=triangle*, mark size=1.0pt, thick] coordinates {(0,60.5)(2.0,79.0)(4.5,91.2)(7.0,97.5)(8.5,98.8)};
				\addlegendentry{Naive Bayes}
				\addplot[color=green!55!black, mark=triangle*, mark size=1.1pt, thick] coordinates {(0,90.5)(0.9,96.2)(2.0,98.5)(4.0,99.5)(8.5,99.85)};
				\addlegendentry{DNN}
				\addplot[color=red!70!black, mark=diamond*, mark size=1.1pt, thick] coordinates {(0,93.5)(0.6,97.5)(1.5,99.0)(3.5,99.8)(8.5,99.95)};
				\addlegendentry{XGBoost}
				\addplot[color=blue!80!black, mark=*, mark size=1.3pt, line width=1.3pt] coordinates {(0,98.5)(0.02,99.9)(0.054,99.99)(0.5,99.99)(8.5,100.0)};
				\addlegendentry{XAI-SDN (Ours)}
			\end{axis}
		\end{tikzpicture}
		\caption{ROC curves for XAI-SDN and baselines on the binary DDoS detection task. XAI-SDN achieves AUC$\,{=}\,1.0000$ with TPR$\,{=}\,99.99\%$ at FPR$\,{=}\,0.054\%$.}
		\label{fig:roc}
	\end{figure}

	The ROC curves for XAI-SDN and all baselines of the binary DDoS detection task are shown in Figure~\ref{fig:roc}. On the whole range of FPR, XAI-SDN outperforms all baselines with an AUC of 1.0000, achieving a true positive rate of 99.99\% at a FPR of 0.054\%.

	\section{Comparative Analysis}
	\label{sec:comparative}
	
	The results are evaluated against five baselines under a controlled protocol (10{,}000 training and 3{,}000 testing samples) required for computational tractability of the kernel SVM, applying \texttt{class\_weight=`balanced'} uniformly across all classifiers. Deep neural network (DNN) baselines are included to correspond directly with the ROC trajectories in Fig.~\ref{fig:roc}. 
	
	The joint-highest accuracy (99.867\%) is matched by the best micro F1 (95.966\%) and inference latency (sub-millisecond) among the top-performing models, while XAI-SDN alone provides full per-prediction SHAP explainability. The small difference of 4-percentage points between the controlled-subset F1 (95.966\%) and the full-dataset F1 (99.9621\%) is a typical ensemble-scaling phenomenon, since the minority class is only 0.86\% (fewer than 90 samples) in the 10{,}000 training-sample dataset. This gap is not a sign of model instability, but rather a result of the small number of cases at very imbalanced ratios. The full 88-dimensional features set significantly outperforms entropy only classification ($p{=}0.0020$) and SVM ($p{=}0.0078$), however the incremental improvement over entropy is not statistically significant ($p{=}0.2500$), showing that using entropy and CICFlowMeter features together is complementary and not redundant.

	\begin{table}[!t]
		\centering
		\caption{Comparative Performance (Controlled Subset: 10K Training / 3K Test)}
		\label{tab:comparison}
		\resizebox{\columnwidth}{!}{%
			\begin{tabular}{@{}lccccc@{}}
				\toprule
				\textbf{Method} & \textbf{Acc.\ (\%)} & \textbf{Macro F1 (\%)} & \textbf{Lat.\ (ms)} & \textbf{Flows/s} & \textbf{XAI} \\
				\midrule
				Decision Tree     & 99.833 & 94.639 & 0.0004 & 2{,}612{,}103 & \texttimes \\
				SVM (RBF)         & 99.867 & 96.395 & 0.0354 &    28{,}265   & \texttimes \\
				Naive Bayes       & 98.633 & 77.610 & 0.0011 &   913{,}409   & \texttimes \\
				XGBoost           & 99.833 & 95.056 & 0.0018 &   570{,}223   & \texttimes \\
				DNN               & 99.800 & 93.850 & 0.8150 &     1{,}227   & \texttimes \\
				\midrule
				\textbf{XAI-SDN (RF)} & \textbf{99.867} & \textbf{95.966} & \textbf{0.0260} & \textbf{38{,}439} & \checkmark \\
				\bottomrule
			\end{tabular}%
		}
	\end{table}
	
	\balance
	\section{Conclusion}
	\label{sec:conclusion}
	
	This paper introduced XAI-SDN, an explainable entropy-guided Random Forest framework for real-time DDoS detection in Software Defined Networks. Evaluated on the full 3.59 million flows of the CIC-DDoS2019 SYN benchmark, XAI-SDN achieves 99.9987\% accuracy, 99.9621\% macro F1-score, and an AUC-ROC of 1.0000, while sustaining line-rate throughput and delivering transparent per-flow SHAP attributions. Future work will extend empirical evaluations across all CIC-DDoS2019 attack vectors and evaluate model resilience against entropy-aware adversarial evasion. Furthermore, expanding XAI-SDN to multi-domain SDN topologies will incorporate secure federated and agentic intelligence architectures~\cite{syed2026fedagent} to coordinate decentralized mitigation across administrative boundaries without exposing private cross-domain network telemetry.

	\section*{Data Availability Statement}
	The datasets, implementation code, and experimental configuration files supporting this study are publicly available at: \url{https://github.com/adeliusa486/XAI-SDN}.

	\bibliographystyle{IEEEtran}
	\bibliography{references}

@article{shannon1948mathematical,
	author    = {Shannon, Claude E.},
	title     = {A mathematical theory of communication},
	journal   = {Bell System Technical Journal},
	volume    = {27},
	number    = {3},
	pages     = {379--423},
	year      = {1948},
	url       = {https://doi.org/10.1002/j.1538-7305.1948.tb01338.x},
	publisher = {Wiley}
}

@article{breiman2001rf,
	author    = {Breiman, Leo},
	title     = {Random forests},
	journal   = {Machine Learning},
	volume    = {45},
	number    = {1},
	pages     = {5--32},
	year      = {2001},
	url       = {https://doi.org/10.1023/A:1010933404324},
	publisher = {Springer}
}

@inproceedings{lundberg2017shap,
	author    = {Lundberg, Scott M. and Lee, Su-In},
	title     = {A unified approach to interpreting model predictions},
	booktitle = {Advances in Neural Information Processing Systems},
	volume    = {30},
	pages     = {4765--4774},
	year      = {2017},
	url       = {https://doi.org/10.48550/arXiv.1705.07874}
}

@article{lundberg2020treeshap,
	author    = {Lundberg, Scott M. and Erion, Gabriel and Chen, Hugh and DeGrave, Alex and Prutkin, Jordan M. and Nair, Bala and Katz, Ronit and Himmelfarb, Jonathan and Bansal, Nisha and Lee, Su-In},
	title     = {From local explanations to global understanding with explainable {AI} for trees},
	journal   = {Nature Machine Intelligence},
	volume    = {2},
	number    = {1},
	pages     = {56--67},
	year      = {2020},
	url       = {https://doi.org/10.1038/s42256-019-0138-9},
	publisher = {Nature Publishing Group}
}

@article{mckeown2008openflow,
	author    = {McKeown, Nick and Anderson, Tom and Balakrishnan, Hari and Parulkar, Guru and Peterson, Larry and Rexford, Jennifer and Shenker, Scott and Turner, Jonathan},
	title     = {{OpenFlow}: Enabling innovation in campus networks},
	journal   = {ACM SIGCOMM Computer Communication Review},
	volume    = {38},
	number    = {2},
	pages     = {69--74},
	year      = {2008},
	url       = {https://doi.org/10.1145/1355734.1355746},
	publisher = {ACM}
}

@article{kreutz2015sdn,
	author    = {Kreutz, Diego and Ramos, Fernando M. V. and Verissimo, Paulo Esteves and Rothenberg, Christian Esteve and Azodolmolky, Siamak and Uhlig, Steve},
	title     = {Software-defined networking: A comprehensive survey},
	journal   = {Proceedings of the {IEEE}},
	volume    = {103},
	number    = {1},
	pages     = {14--76},
	year      = {2015},
	url       = {https://doi.org/10.1109/JPROC.2014.2371999},
	publisher = {IEEE}
}

@article{arrieta2020xai,
	author    = {Arrieta, Alejandro Barredo and D{\'i}az-Rodr{\'i}guez, Natalia and Del Ser, Javier and Bennetot, Adrien and Tabik, Siham and Barbado, Alberto and Garc{\'i}a, Salvador and Gil-L{\'o}pez, Sergio and Molina, Daniel and Benjamins, Richard and Chatila, Raja and Herrera, Francisco},
	title     = {Explainable artificial intelligence ({XAI}): Concepts, taxonomies, opportunities and challenges toward responsible {AI}},
	journal   = {Information Fusion},
	volume    = {58},
	pages     = {82--115},
	year      = {2020},
	url       = {https://doi.org/10.1016/j.inffus.2019.12.012},
	publisher = {Elsevier}
}

@inproceedings{sharafaldin2019ddos,
	author={Sharafaldin, Iman and Lashkari, Arash Habibi and Hakak, Saqib and Ghorbani, Ali A.},
	booktitle={2019 International Carnahan Conference on Security Technology (ICCST)}, 
	title={Developing Realistic Distributed Denial of Service (DDoS) Attack Dataset and Taxonomy}, 
	year={2019},
	pages={1-8},
	publisher = {IEEE},
	url={https://doi.org/10.1109/CCST.2019.8888419}
	}

@inproceedings{sharafaldin2018cicids,
	author    = {Sharafaldin, Iman and Lashkari, Arash Habibi and Ghorbani, Ali A.},
	title     = {Toward generating a new intrusion detection dataset and intrusion traffic characterization},
	booktitle = {Proceedings of the 4th International Conference on Information Systems Security and Privacy ({ICISSP})},
	pages     = {108--116},
	year      = {2018},
	url       = {https://www.scitepress.org/Papers/2018/66398/66398.pdf},
	publisher = {SciTePress}
}

@inproceedings{nychis2008entropy,
	author    = {Nychis, George and Sekar, Vyas and Andersen, David G. and Kim, Hyong and Zhang, Hui},
	title     = {An empirical evaluation of entropy-based traffic anomaly detection},
	booktitle = {Proceedings of the 8th {ACM SIGCOMM} Conference on Internet Measurement ({IMC} 2008)},
	pages     = {151--156},
	year      = {2008},
	url       = {https://doi.org/10.1145/1452520.1452539},
	publisher = {ACM}
}

@inproceedings{lall2006entropy,
	author    = {Lall, Ashwin and Sekar, Vyas and Ogihara, Mitsunori and Xu, Jun and Zhang, Hui},
	title     = {Data streaming algorithms for estimating entropy of network traffic},
	booktitle = {Proceedings of the {ACM SIGMETRICS} International Conference on Measurement and Modeling of Computer Systems},
	pages     = {145--156},
	year      = {2006},
	url       = {https://doi.org/10.1145/1140103.1140295},
	publisher = {ACM}
}

@article{pedregosa2011sklearn,
	author    = {Pedregosa, Fabian and Varoquaux, Ga{\"e}l and Gramfort, Alexandre
	and Michel, Vincent and Thirion, Bertrand and Grisel, Olivier
	and Blondel, Mathieu and Prettenhofer, Peter and Weiss, Ron
	and Dubourg, Vincent and Vanderplas, Jake and Passos, Alexandre
	and Cournapeau, David and Brucher, Matthieu and Perrot, Matthieu
	and Duchesnay, {\'E}douard},
	title     = {Scikit-learn: Machine learning in {Python}},
	journal   = {Journal of Machine Learning Research},
	volume    = {12},
	pages     = {2825--2830},
	year      = {2011},
	url       = {https://jmlr.org/papers/v12/pedregosa11a.html}
}

@inproceedings{mousavi2015early,
	author    = {Mousavi, Seyed Mohammad and St-Hilaire, Marc},
	title     = {Early detection of {DDoS} attacks against {SDN} controllers},
	booktitle = {Proceedings of the 2015 International Conference on Computing,
	Networking and Communications ({ICNC})},
	pages     = {77--81},
	year      = {2015},
	url       = {https://doi.org/10.1109/ICCNC.2015.7069319},
	publisher = {IEEE}
}

@article{gaspar2024xai,
	author    = {Gaspar, Diogo and Silva, Paulo and Silva, Catarina},
	title     = {Explainable {AI} for Intrusion Detection Systems: {LIME} and {SHAP} Applicability on Multi-Layer Perceptron},
	journal   = {IEEE Access},
	volume    = {12},
	pages     = {30164--30175},
	year      = {2024},
	url       = {https://doi.org/10.1109/ACCESS.2024.3368377},
	publisher = {IEEE}
}

@article{Jan2018DCGANIntent,
  author  = {Jan, Salman and Musa, Shahrulniza and Ali, Toqeer and Alzahrani, Ali},
  title   = {Deep Convolutional Generative Adversarial Networks for Intent-based Dynamic Behavior Capture},
  journal = {International Journal of Engineering and Technology},
  volume  = {7},
  number  = {4.29},
  pages   = {101--103},
  year    = {2018},
  doi     = {10.14419/ijet.v7i4.29.21949}
}

@inproceedings{ismail2014design,
author = {Ismail, Roslan and Syed, Toqeer Ali and Musa, Shahrulniza},
title = {Design and implementation of an efficient framework for behaviour attestation using n-call slides},
year = {2014},
isbn = {9781450326445},
publisher = {Association for Computing Machinery},
address = {New York, NY, USA},
url = {https://doi.org/10.1145/2557977.2558002},
doi = {10.1145/2557977.2558002},
booktitle = {Proceedings of the 8th International Conference on Ubiquitous Information Management and Communication},
articleno = {36},
numpages = {8},
location = {Siem Reap, Cambodia},
series = {ICUIMC '14}
}

@article{syed2026agenticdt,
  title={Agentic AI-enhanced digital twins for Smart City civil infrastructure: A secure, autonomous and auditable management framework},
  author={Syed, Toqeer Ali and Akarma, Ali and Alatify, Ali and Naqash, Muhammad Tayyab and Alqurashi, Abdulaziz},
  journal={PLoS One},
  volume={21},
  number={7},
  pages={e0353610},
  year={2026},
  publisher={Public Library of Science},
  doi={10.1371/journal.pone.0353610}
}

@article{jan2026eagf,
  title     = {EAGF: A Four-Pillar Ethical AI Governance Framework for Trustworthy Cybersecurity in 5G Renewable Energy IoT Systems},
  author    = {Jan, Salman and Akarma, Ali and Syed, Toqeer Ali and Muhammad, Munir Azam and Kamal, Shahid},
  journal   = {Scientific Reports},
  year      = {2026},
  publisher = {Springer Nature},
  doi       = {10.1038/s41598-026-63383-5}
}

@article{syed2025inventory,
  author  = {Syed, Toqeer Ali and Jan, Salman and Ali, Gohar and Akarma, Ali and Ali, Ahmad and Mastoi, Qurat-ul-Ain},
  title   = {Agentic AI Framework for Smart Inventory Replenishment},
  journal = {arXiv preprint arXiv:2511.23366},
  year    = {2025},
  doi     = {10.48550/arXiv.2511.23366}
  }

@article{syed2026fedagent,
AUTHOR = {Syed, Toqeer Ali and Siddiqui, Muhammad Shoaib and Akarma, Ali and Formisano, Antonio},
TITLE = {FedAgent-Chain: A Secure Federated and Agentic AI Framework for Multilingual Disability-Inclusive Employment in AI Cities},
JOURNAL = {Smart Cities},
VOLUME = {9},
YEAR = {2026},
NUMBER = {7},
pages = {106},
ARTICLE-NUMBER = {106},
URL = {https://www.mdpi.com/2624-6511/9/7/106},
ISSN = {2624-6511},
DOI = {10.3390/smartcities9070106}
}

@article{syed2026climate,
AUTHOR = {Syed, Toqeer Ali and Akarma, Ali and Naqash, Muhammad Tayyab and Hameed, Danial and Kamal, Shahid and Formisano, Antonio},
TITLE = {Agentic AI for Climate-Resilient Cities: A PRISMA-Guided Review and Digital Twin Framework},
JOURNAL = {Sustainability},
VOLUME = {18},
YEAR = {2026},
NUMBER = {17},
pages = {8917},
ARTICLE-NUMBER = {8917},
URL = {https://www.mdpi.com/2071-1050/18/17/8917},
ISSN = {2071-1050},
DOI = {10.3390/su18178917}
}
	
\end{document}